\documentclass[prb,amssymb,superscriptaddress,floatfix,a4paper,twocolumn]{revtex4}
\usepackage{graphicx}
\usepackage{bm}
\usepackage{amsmath}

\begin{document}

\title{Domain growth within the backbone of the three-dimensional $\pm J$ Edwards-Anderson spin glass}

\author{F. Rom\'a}
\affiliation{CONICET, Departamento de F\'{\i}sica, Universidad
Nacional de San Luis, D5700HHW San Luis, Argentina}

\author{S. Bustingorry}
\affiliation{CONICET, Centro At{\'{o}}mico Bariloche, 8400 San
Carlos de Bariloche, R\'{\i}o Negro, Argentina}

\author{P. M. Gleiser}
\affiliation{CONICET, Centro At{\'{o}}mico Bariloche, 8400 San
Carlos de Bariloche, R\'{\i}o Negro, Argentina}

\date{\today}

\begin{abstract}

The goal of this work is to show that a ferromagnetic-like domain growth process takes place within the backbone of the three-dimensional $\pm J$ Edwards-Anderson (EA) spin glass model. To sustain this affirmation we study the heterogeneities displayed in the out-of-equilibrium dynamics of the model. We show that both correlation function and mean flipping time distribution present features that have a direct relation with spatial heterogeneities, and that they can be characterized by the backbone structure. In order to gain intuition we analyze the pure ferromagnetic Ising model, where we show the presence of dynamical heterogeneities in the mean flipping time distribution that are directly associated to ferromagnetic growing domains. We extend a method devised to detect domain walls in the Ising model to carry out a similar analysis in the three-dimensional EA spin glass model. This allows us to show that there exists a domain growth process within the backbone of this model.

\end{abstract}


\maketitle

\section{Introduction}
\label{sec:intro}

From a first approximation, glasses are characterized by two main features: a spatially disordered structure resembling the high temperature state and a slow dynamics with the typical equilibration time scale exceeding the accessible laboratory time scale. It has been a matter of intensive study during the past decades what are the microscopic mechanisms behind the nature of the low temperature glass phase. In order to do that different methods and techniques have been used from both experimental and theoretical approaches. In particular, during the last years the concept of heterogeneity has been increasingly used to study the slow relaxation of glassy systems, both for structural and spin glasses. For structural glasses, this has led to the identification of slow and fast groups of particles which are spatially correlated and evolving with time.~\cite{weeks2000,appignanesi2006,Castillo,appignanesi2009} For spin glasses, dynamical and spatial heterogeneities have been intensively studied, both from a coarse grained perspective,~\cite{Chamon02,Castillo02,Castillo03,Chamon04} where the dynamics is analyzed within a time or space window, or from single spin analysis,~\cite{Montanari03a,Montanari03b} where each spin is analyzed independently for a given disorder realization.

In particular, the ground state (GS) topology of the three-dimensional (3D) $\pm J$ Edwards-Anderson (EA) spin glass model~\cite{EA} presents intrinsic spatial heterogeneities. For a given disorder realization, the multiple-degenerated GS configurations can be used to identify a constrained structure, usually referred as the ``backbone''.~\cite{Vannimenus,Barahona} We have recently incorporated this spatial heterogeneous character of the GS into the analysis of dynamical heterogeneities present in the out of equilibrium dynamics~\cite{Roma2,Roma3} and also in the study of equilibrium properties of the system.~\cite{rubio2009} In Ref.~\onlinecite{Roma2} we have analyzed the two-dimensional EA model and we have correlated the presence of slow and fast sets of spins with the backbone structure. Besides, for the 3D EA model we have analyzed the out-of-equilibrium violation of the fluctuation-dissipation theorem. The results suggest that a domain growth process might be present within the backbone, while those spins outside the backbone tend to equilibrium behavior.~\cite{Roma3} Also, by using the damage-spreading technique, we have results showing that ferromagnetic-like order grows in the 3D EA model below the pure ferromagnetic critical temperature $T_c$, pointing to the presence of a Griffiths-like phase in the range $T_{\mathrm{g}} < T < T_{\mathrm{c}}$,~\cite{rubio2009} where $T_{\mathrm{g}}$ is the glass transition temperature. Again, these results are suggestive of a domain growth process inside the backbone structure.

In the present work we further pursue the origin of the domain growth signatures found in the $\pm J$ EA model by using exhaustive numerical simulations. With this in mind we analyze the out of equilibrium dynamics in the 3D case while using information of the GS topology as a main input. By directly comparing with the dynamics of the pure ferromagnetic Ising model we are able to show strong evidence supporting the presence of a ferromagnetic-like domain growth process within the largest cluster of the backbone for $T<T_{\mathrm{g}}$.

\subsection*{Outline of the paper}
\label{sec:outline}

Here, we outline the paper in order to guide the reader through the presentation of the main results. Section~\ref{sec:model} presents the model and the main characteristics of its GS topology. This includes the important definition of the backbone, allowing to classify the spins based on its solidary and non-solidary character. This is a useful notion to characterize spatial heterogeneities and is used throughout the paper. In Sec.~\ref{sec:corr} we show how these spatial heterogeneities are present in the two-times correlation function, establishing a direct link with typical finite-temperature dynamical heterogeneities. Then, the strong heterogeneous character of the mean flipping time distribution is analyzed in Sec.~\ref{sec:dist}.

Since the dynamics within the set of solidary spins prompt to an analysis in terms of domain growth, we compare it with the well know pure ferromagnetic Ising model. Then, in Sec.~\ref{sec:ising} we present results for the out-of-equilibrium mean flipping time distribution of the Ising model and its correlation with the domain growth process which takes place.

In Sec.~\ref{sec:trans} we describe, through a gauge transformation, how ferromagnetic-like order can be sustained within the backbone structure of the 3D $\pm J$ EA model. It is shown that the percolative set of solidary spins can be thought of as a ferromagnetic-like system. All the information is gathered in Sec.~\ref{sec:growth}, where it is shown, after a comparison of the mean flipping time distribution of the EA model with the Ising model, that a ferromagnetic-like structure is growing inside the backbone of the 3D EA model. Section~\ref{sec:conc} is devoted to the discussion and conclusions. Finally, in the Appendix we characterize finite size effects by analyzing the out-of-equilibrium correlation function.

\section{Model}
\label{sec:model}

We consider the 3D $\pm J$ EA spin glass model,~\cite{EA} which consists of a set of $N$ Ising spins $\sigma_i = \pm 1$ placed in a cubic lattice of linear dimension $L$, with periodic boundary conditions in all directions. The Hamiltonian is
\begin{equation}
H = \sum_{( i,j )} J_{ij} \sigma_{i} \sigma_{j},
\label{hamiltonian}
\end{equation}
where $( i,j )$ indicates a sum over the six nearest neighbors. The coupling constants or bonds, $J_{ij}$'s, are independent random variables drawn from a bimodal distribution, i.e. $J_{ij}=\pm 1$ with equal probability.

We shall focus on the out-of-equilibrium dynamics of the model. The initial condition, $t=0$, corresponds to a completely disordered spin configuration, 
that is to $\sigma_i = \pm1$ randomly chosen, which mimics a quench from $T=\infty$. The working temperature is set to $T<T_g=1.12$.~\cite{Katzgraber2006} Using a two-times protocol, after a waiting time $t_{\text{w}}$ we measure different time correlation functions, which indeed depend on the two times, $t$ and $t_{\text{w}}$. We use standard Glauber dynamics with sequential random updates using a continuous time Monte Carlo algorithm.~\cite{Bortz1975}

The EA spin glass model has a highly degenerate GS. From all the GS configurations corresponding to a single disorder realization it is possible to identify a constrained structure, which is called the rigid lattice.~\cite{Barahona,Vannimenus} This structure is composed of those bonds which do not change its state --satisfied or frustrated-- in all GS configurations. Using this information one can divide the spins of the system in two main sets: \textit{solidary} spins, defined as those connected through the rigid lattice and thus maintaining their relative orientation in all GS configurations, and \textit{non-solidary} spins, which are simply the complementary set. Both, the rigid lattice and the set of solidary spins compose the backbone of the system.

The topological properties of the backbone have been intensively studied recently.~\cite{Roma2006,Roma2009a} Using a parallel tempering Monte Carlo algorithm it is possible to systematically arrive at GS configurations,~\cite{Roma2009b} necessary to determine the backbone structure. We stress that the determination of backbone structures for each disorder realization requires to visit $O(N)$ times the GS. In two dimensions the backbone does not percolate, which is consistent with the absence of a finite glass transition temperature. On the other hand, in three dimensions this constrained structure percolates and comprises $76\%$ of the of the spins. Through extensive numerical simulations we were able to obtain the backbone structure for $M=1000$ different sample, i.e. different realizations of bond disorder, with $N=8^3$ spins. This allows us to measure quantities that are averaged over many disorder configurations and thus to present results whose interpretation has a general physical meaning and is not restricted to a particular realization. In order to determine finite size effects we also compare the results obtained with $L=8$ with larger system sizes when possible, and restrict our analysis to the parameter region where these effects are negligible.

\section{Two-times correlation function}
\label{sec:corr}

\begin{figure}[!tbp]
\includegraphics[width=7cm,clip=true]{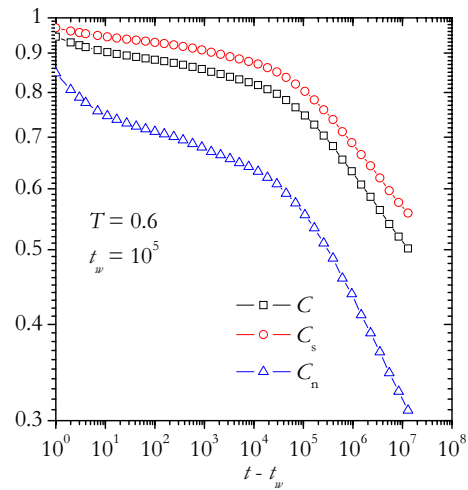}
\caption{\label{figure1} (Color online) Correlation functions for $L=8$ system at $T=0.6$. The figure shows the full correlation function (black squares), the solidary (red circles) and non-solidary (blue triangles) contribution. The average were carried out over $M=10^3$ samples with $m=10$ thermal histories for each sample.  Curves correspond to $t_{\text{w}}=10^5$.}
\end{figure}

\begin{figure}[!tbp]
\includegraphics[width=7cm,clip=true]{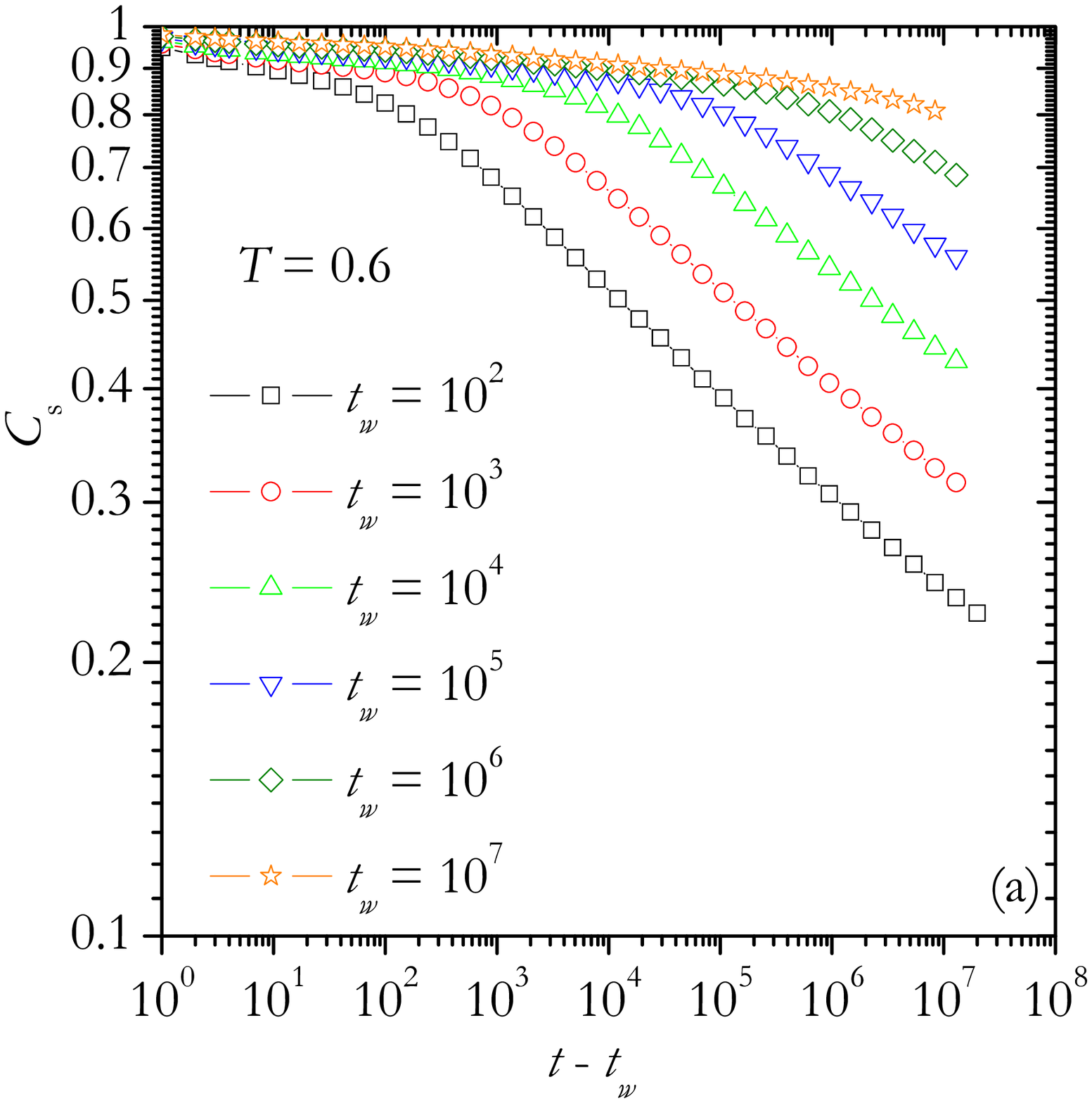}
\includegraphics[width=7cm,clip=true]{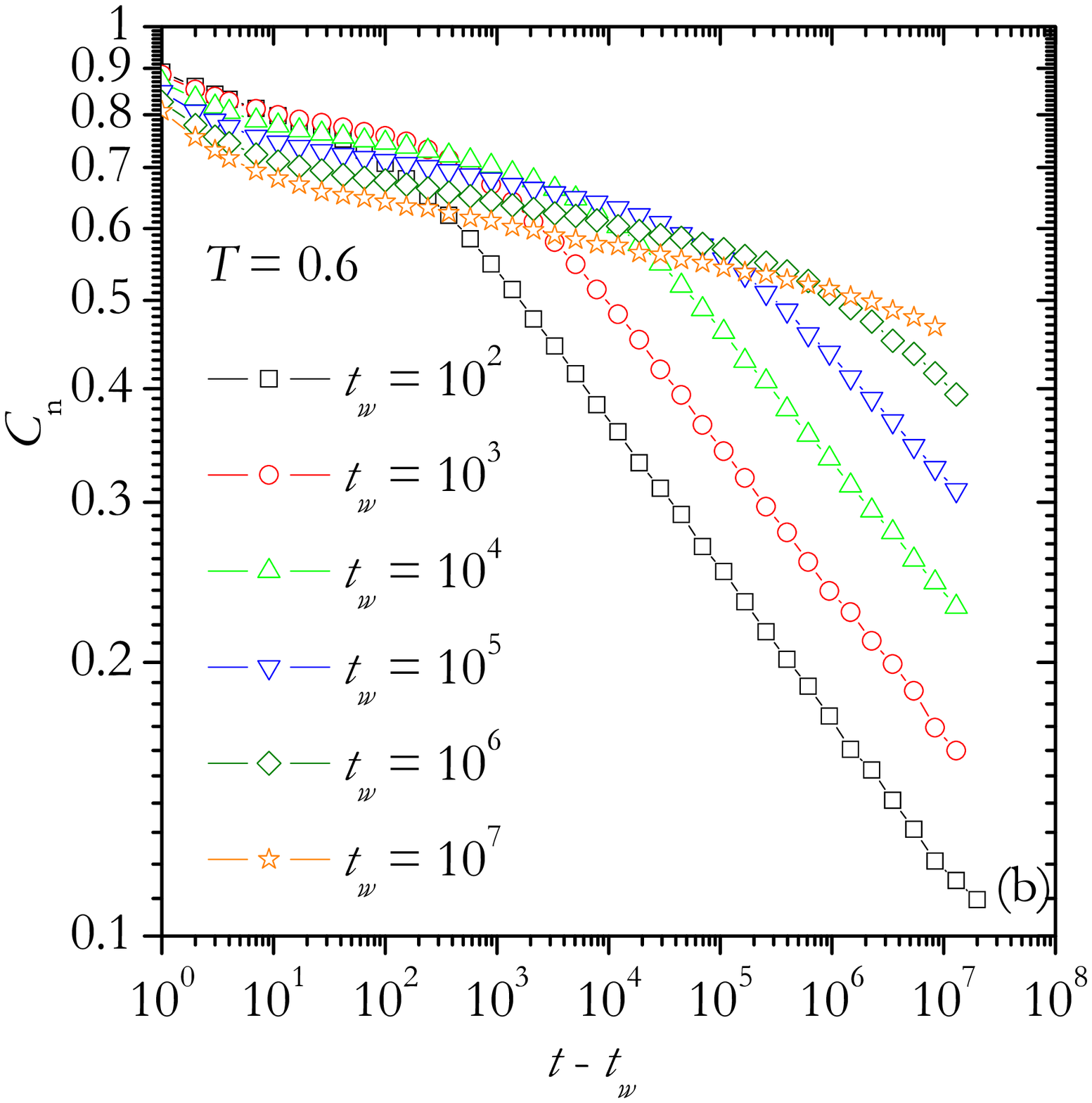}
\caption{\label{figure2} (Color online) Correlation functions restricted over the sets of (a) solidary and (b) non-solidary spins. Curves corresponds to $L=8$ at $T=0.6$ and for different values of $t_{\text{w}}$ as indicated. The average were carried out over $M=10^3$ samples with $m=10$ thermal histories for each sample.}
\end{figure}

We shall first focus on the spin two-times correlation function defined for $t>t_{\text{w}}$ as
\begin{equation}
C (t,t_{\text{w}}) = \left [ \left \langle \frac{1}{N} \sum_{i=1}^N
\sigma_i^{\alpha}(t) \sigma_i^{\alpha}(t_{\text{w}}) \right \rangle \right
]_{\text{av}}, \label{corr}
\end{equation}
where $\sigma_i^{\alpha}$ are Ising spins of sample $\alpha$, $\langle \dots \rangle$ is an average over $m$ thermal histories, that is, over different initial conditions and realizations of the thermal noise, and $[\dots]_{\text{av}}$ indicates average over $M$ samples. For clarity, we keep the $\alpha$ index for single disorder realizations. In the Appendix we discuss finite size effects and we present evidence allowing as to restrict the analysis of two-times quantities to systems with $L=8$.

In order to illustrate how GS information is reflected in the finite temperature dynamics we show here the non-trivial separation of the two-times correlation function. While the system is evolving as a whole, one can compute the correlation function restricting the sum over a fraction of the whole sample. Indeed, the correlation function restricted over the sets of solidary and non-solidary spins are, respectively,
\begin{equation}
C_\text{s}(t,t_{\text{w}}) = \left [ \left \langle
\frac{1}{N_\text{s}^{\alpha}} \sum_{\text{s}}
\sigma_i^{\alpha}(t) \sigma_i^{\alpha}(t_{\text{w}}) \right \rangle \right
]_{\text{av}}, \label{corr_s}
\end{equation}
and
\begin{equation}
C_\text{n}(t,t_{\text{w}}) = \left [ \left \langle
\frac{1}{N_\text{n}^{\alpha}} \sum_{\text{n}}
\sigma_i^{\alpha}(t) \sigma_i^{\alpha}(t_{\text{w}}) \right \rangle \right
]_{\text{av}}. \label{corr_n}
\end{equation}
In these definitions $N_\text{s}^{\alpha}$ and $N_\text{n}^{\alpha}$ are, respectively, the number of solidary and non-solidary spins of sample $\alpha$, and the sums are restricted to the corresponding regions of each sample. Figure~\ref{figure1} presents the correlation function for $T=0.6$ and $t_{\text{w}}=10^5$, comparing the full correlation function (black squares) with the correlation function restricted to the solidary (red circles) and non-solidary (blue triangles) spins. As discussed in Ref.~\onlinecite{Roma2} for the two-dimensional case, the separation is non trivial. For short times the solidary spins are highly correlated presenting a very slow decay. At longer times a second regime with a faster decay appears. Thus, those spins which are highly correlated in the GS configurations, tend to maintain their correlation in time. On the other hand, non-solidary spins present a rapid decay of the correlation at short times, followed by a second faster decay. From this separation one can relate solidary and non-solidary spins to groups of spins with slow and fast dynamics, which is characteristic of dynamical heterogeneities. Therefore, the results show that the spatial heterogeneity appearing in the GS configuration is correlated with the dynamical heterogeneity of the out-of-equilibrium finite low-temperature dynamics.

Figures~\ref{figure2}(a) and (b) show, respectively, the correlation functions of solidary and non-solidary spins for a system with $L=8$ at $T=0.6$ and different waiting times. One can observe that the correlation function of solidary spins behaves qualitatively similar to the full correlation function shown in Fig.~\ref{figure1}(a). Note also, that for solidary spins, at longer waiting times the initial decay is smaller. This means that for longer waiting times solidary spins are more correlated. On the contrary, the correlation function corresponding to non-solidary spins, Fig.~\ref{figure2}(b), behaves differently: the larger the values of the waiting time, the more pronounced is the first decay, indicating that non-solidary spins become less and less correlated with time.

From the analysis of the correlation function given above, we expect that the time scale separation already observed in the preasymptotic regime of the two-dimensional EA spin glass model~\cite{Roma2} will also be present in three dimensions. Also, it is worth stressing that in contrast to the two dimensional case, in 3D the backbone structure percolates through the sample,~\cite{Roma2009a} and is therefore potentially capable of sustaining some kind of phase. In the next section we analyze the time scale separation through the analysis of the mean flipping time distribution.

\section{Flipping time distributions}
\label{sec:dist}

\begin{figure}[!tbp]
\includegraphics[width=7cm,clip=true]{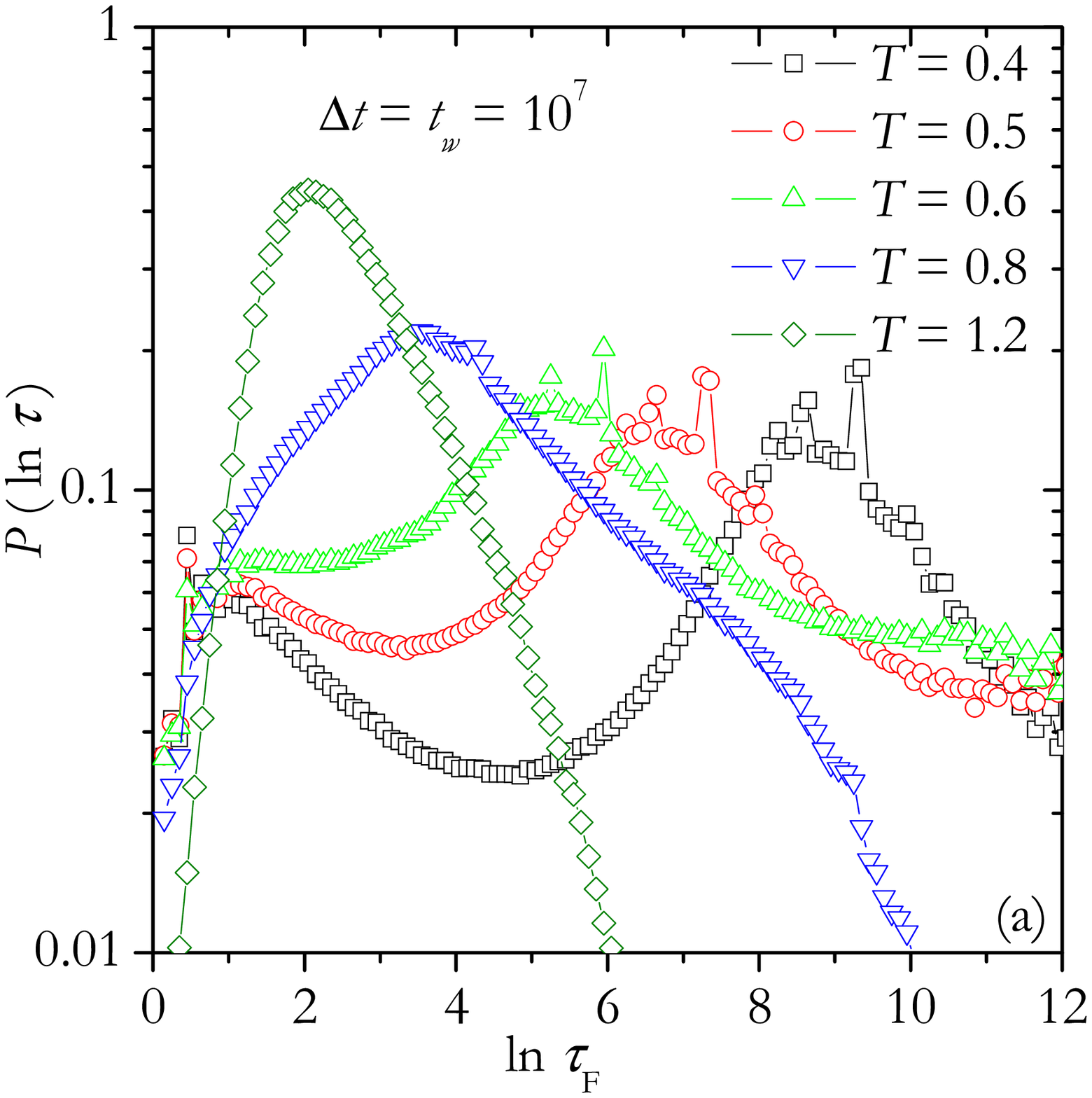}
\includegraphics[width=7cm,clip=true]{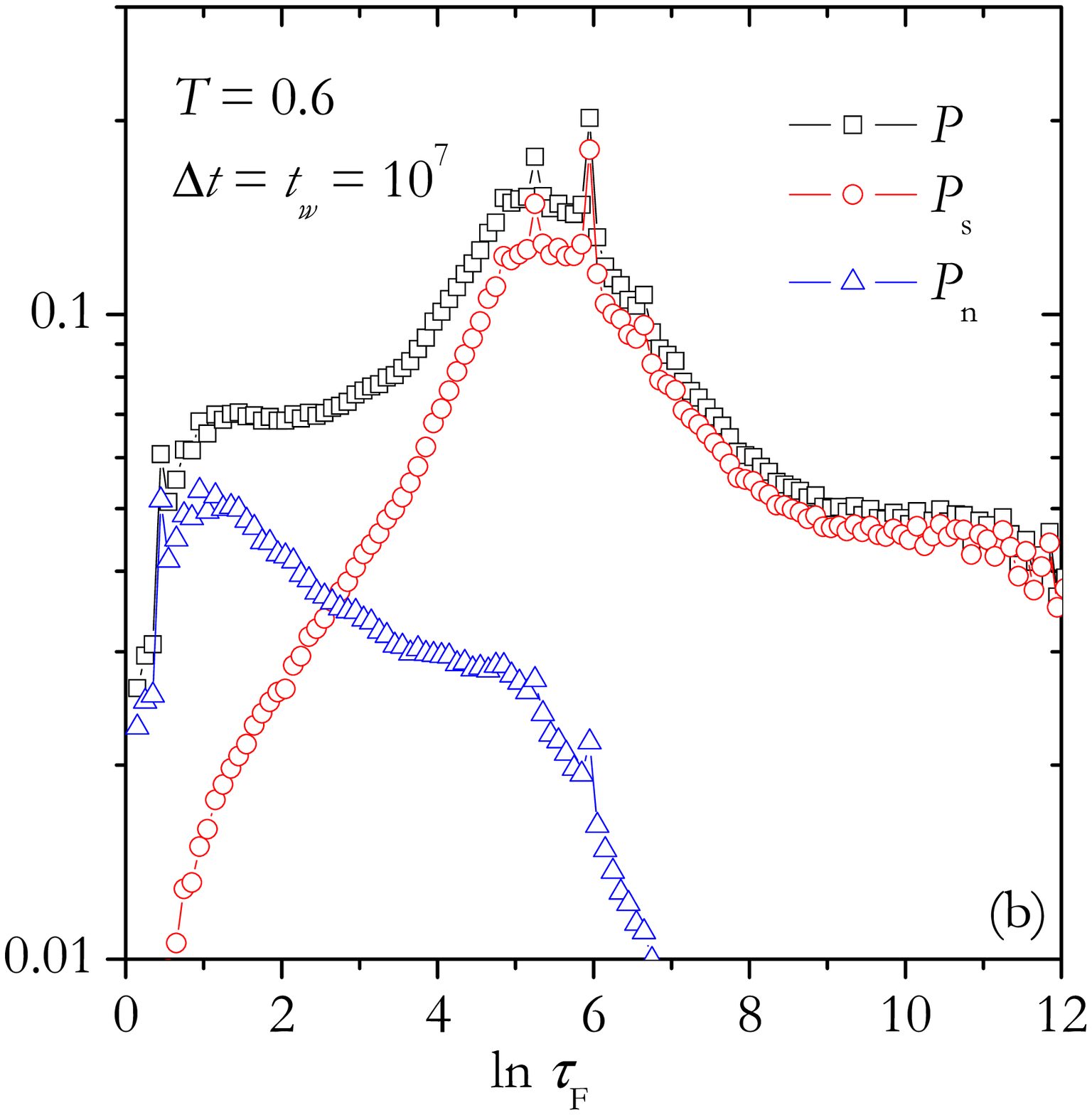}
\caption{\label{figure3} (Color online) Mean flipping time distribution $P$ as function of $\ln \tau_{\text{F}}$, for samples of size $L=8$ for $\Delta t=t_{\text{w}}=10^7$. Panel (a) presents the temperature dependence of the full $P$. Panel (b) shows for $T=0.6$ the two contributions, $P_\text{s}$ and $P_\text{n}$, corresponding to solidary and non-solidary spins. The average were carried out over $M=10^3$ samples, with $m=10$ thermal histories for each sample.}
\end{figure}

\begin{figure*}[!tbp]
\includegraphics[width=5.5cm,clip=true]{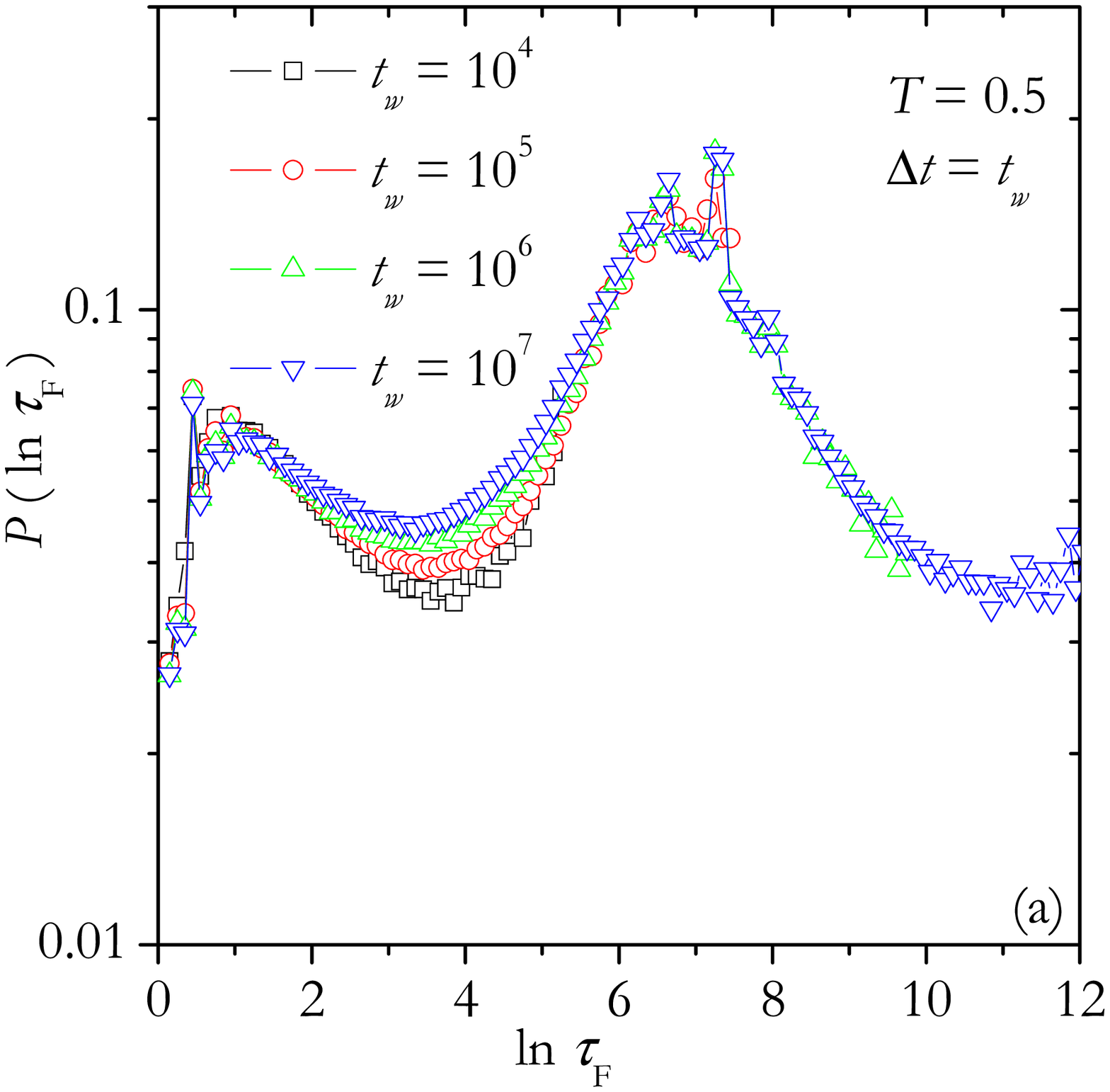}
\includegraphics[width=5.5cm,clip=true]{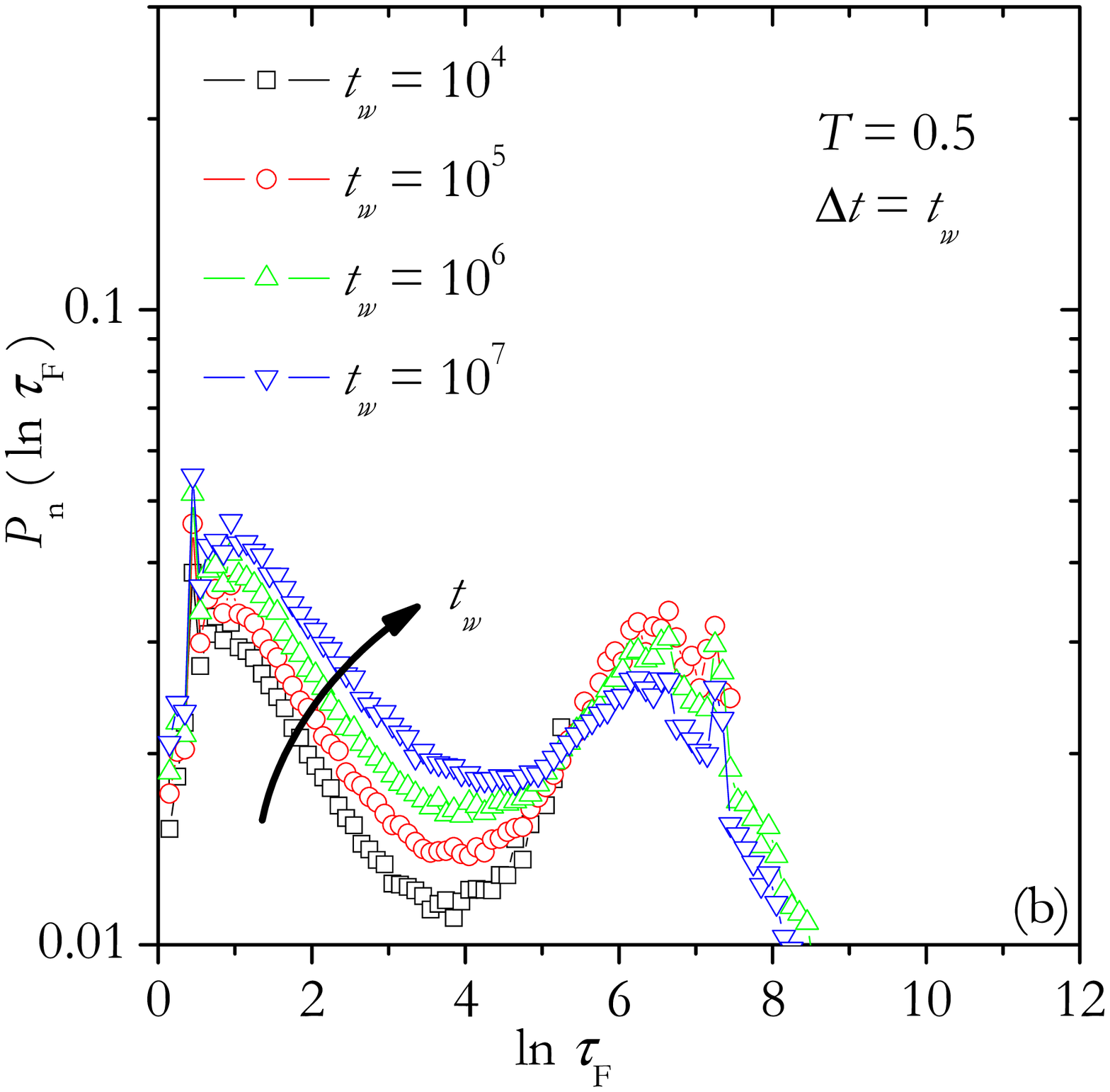}
\includegraphics[width=5.5cm,clip=true]{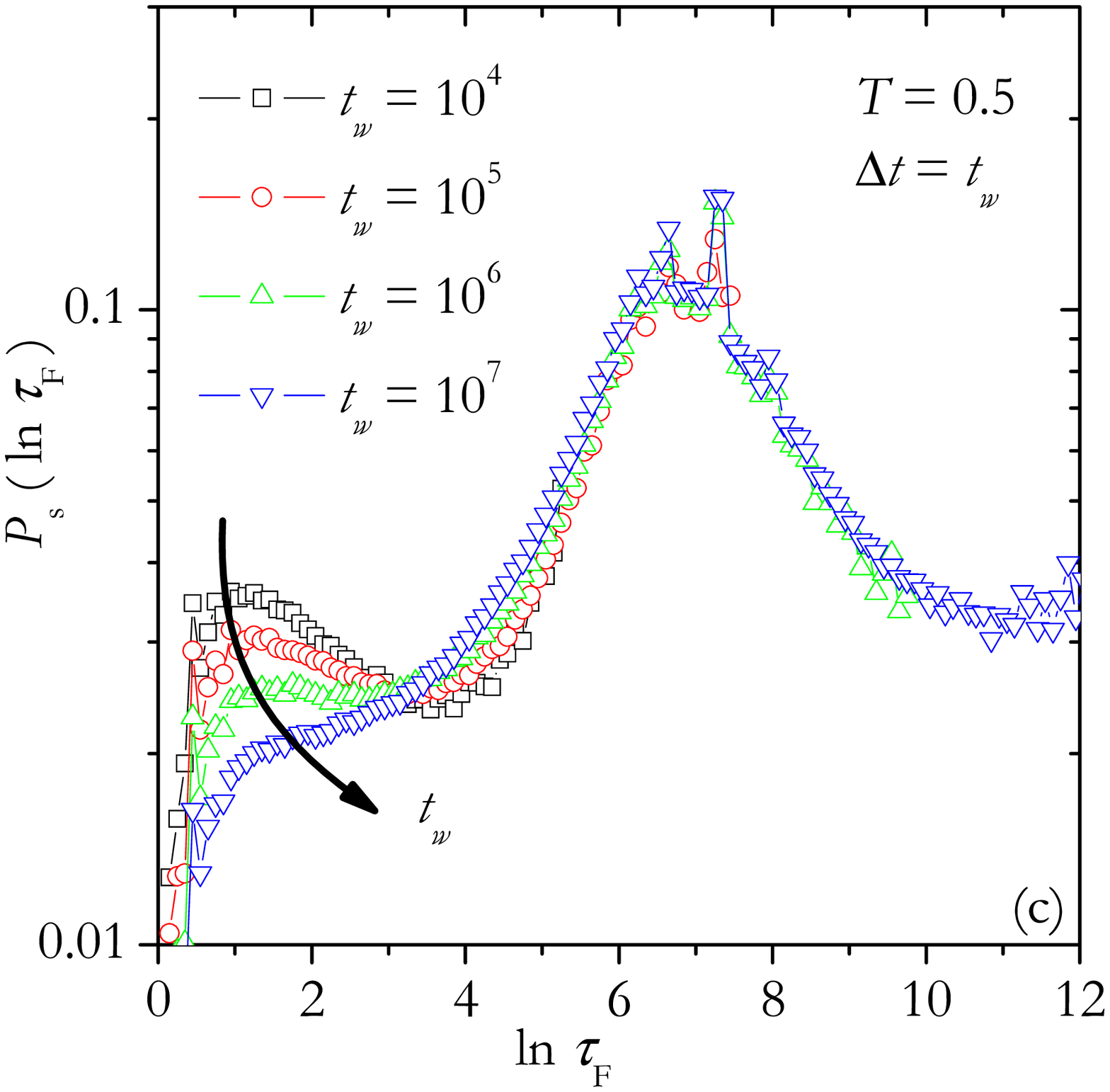}
\caption{\label{figure4} (Color online) The mean flipping time probability distribution at $T=0.5$ for different times and waiting times as indicated, keeping $\Delta t = t_{\text{w}}$. Different panels correspond to: (a) all spins, $P$, (b) non-solidary spins, $P_\text{n}$, and (c) solidary spins, $P_\text{s}$. The average were carried out over $M=10^3$ and $m=10$ for $L=8$. }
\end{figure*}

In this section we shall exploit the analysis of the mean flipping time distribution as an effective way to unveil time scale separations. In each sample we measure the number of flips $(N_{\text{F}})$ done by every spin within a time window extending from $t_{\text{w}}$ to $t$, $\Delta t = t-t_{\text{w}}$. The mean flipping time $\tau_{\text{F}}$ for a given $t_{\text{w}}$ and $t$ is defined as the time window size divided by the number of flips: $\tau_{\text{F}}(t,t_{\text{w}})= \Delta t/ N_{\text{F}}$.~\cite{Ricci} We measure the mean flipping time distribution for each sample and then we average the distribution over disorder realizations to obtain $P(\tau_{\text{F}})=[\langle P^\alpha(\tau_{\text{F}}) \rangle]_{\text{av}}$.~\cite{Ricci} Due to the broadness of the distribution we use a logarithmic scale for the argument, i.e. $P(\ln \tau_{\text{F}})$. Figure~\ref{figure3}(a) shows $P$ as function of $\ln \tau_{\text{F}}$, for samples of size $L=8$ and different temperatures. We show here the particular case of $\Delta t=t_{\text{w}}=10^7$ and present the time evolution below.

The distribution corresponding to $T=1.2$ shows a single peak at small $\tau_{\text{F}}$, corresponding to a more homogeneous high temperature situation. As temperature is decreased the mean flipping time distribution becomes broader, and eventually develops a bimodal shape, with a clear separation of time scales. This  strong dynamical heterogeneity was shown for the first time by Ricci-Tersenghi and Zecchina.~\cite{Ricci} It is also worth stressing that besides the clear bimodal shape, the mean flipping time distribution presents an internal structure characterized by several small and sharp peaks. These peaks, that are also present in Ref.~\onlinecite{Ricci}, follow the general temperature dependence of $P$. However, their origin is not clear, and will be left as an open question for future work.

Now, using the information of the GS for $L=8$, we show in Fig.~\ref{figure3}(b) the mean flipping time probability distribution constrained to the sets of solidary and non-solidary spins, $P_\text{s}$ and $P_\text{n}$, respectively. As was already observed for the two-dimensional case,~\cite{Roma2} the main contribution of $P_\text{s}$ and $P_\text{n}$ are to the slow and fast peaks of the full $P$. 

Although we conclude from the study of the two-times correlation function that solidary spins are more correlated in time, this does not imply that they are always slow. Indeed, a solidary spin of a given sample does not always necessarily contribute to the slow right peak. This is also true for non-solidary spins, which can contribute to both, fast and slow peaks of the mean flipping time distribution functions. This is more evident if one follows the evolution of the mean flipping time distribution, which we present in Fig.~\ref{figure4} for $T=0.5$. For clarity, we only show data for $\Delta t = t_{\text{w}}$, the results being qualitatively similar for other values. Figure~\ref{figure4}(a) shows the full $P$, where it can be observed that the bimodal shape is conserved and the valley between the two peaks is deeper at shorter times. In Figs.~\ref{figure4}(b) and (c) we show the non-solidary and solidary contributions, respectively, to the mean flipping time distribution. Non-solidary spins clearly present a second peak related to slow flips, which indicates, as expected, that the set of non-solidary spins is not completely equivalent to a high temperature paramagnetic phase, which would be composed only of fast spins. Besides, $P_\text{n}$ slowly varies with time, the first peak and the valley between peaks being higher for longer times. In Fig.~\ref{figure4}(c) one can observe that $P_\text{s}$ also presents two peaks at short times. However, whilst the second peak does not change with time, the first peak presents a drastic fall. This is a strong and important difference between the relaxational behavior of solidary and non-solidary spins. A clue to understand this behavior comes again from the information contained in the GS configurations. Before going into more detail we will analyze a canonical model in statistical physics: the Ising model. This will allow us to gain intuition and also to determine the key observables necessary for a quantitative analysis. 

\section{The Ising model}
\label{sec:ising}

\begin{figure}[!tbp]
\includegraphics[width=7cm,clip=true]{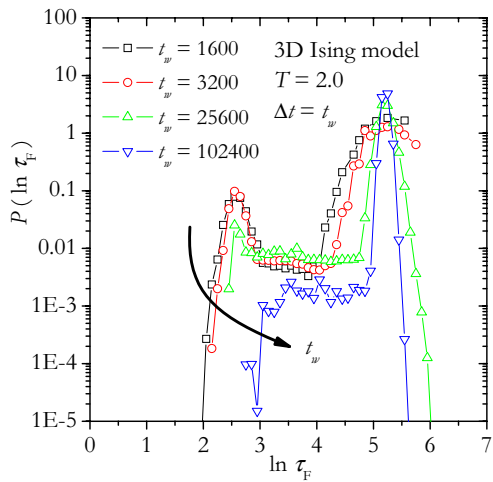}
\caption{\label{figure5} (Color online) The mean flipping time distribution of the 3D Ising model at $T=2.0$ for different values of $t_{\text{w}}$ with $\Delta t = t_{\text{w}}$. The average were carried out over $m=5 \times 10^4$ thermal histories for $L=10$.}
\end{figure}

In this section we use the same protocol defined in the previous section to analyze the dynamics of the pure 3D ferromagnetic Ising model, where all bonds are identical, $J_{ij}=-J$. In particular, we simulate a quench from $T=\infty$ to a working temperature $T<T_c=4.515$.~\cite{binder2001}

In Fig. ~\ref{figure5} we present the time evolution of $P(\ln \tau_{\text{F}})$ when the system is quenched to $T=2.0$. We use here $L=10$ in order to have a system size comparable to the one used for the EA model. We also used $L=50$ and obtained the same qualitative behaviour. For short times the distribution presents a clear bimodal shape. However, for increasing $t_{\text{w}}$, the behavior is different from the one observed in the full mean flipping time distribution of the EA model, Fig.~\ref{figure4}(a). In particular, the peak on the left rapidly falls with increasing waiting times. This behavior strongly resembles the one observed for solidary spins in Fig.~\ref{figure4}(c).

The evolution of the Ising model below its critical temperature serves as a canonical example of growing ferromagnetic order. In order to advance further in understanding the behavior of the mean flipping time distribution we will present a quantitative characterization of the domain wall dynamics. This can be done by implementing the algorithm of Hinrichsen and Antoni,~\cite{Hinrichsen1998} which permits to identify for each of the thermal histories which spins flip due to a domain wall. The main idea of the algorithm is to simulate simultaneously three replicas of a system with different initial conditions (disordered, all spins up, and all spins down) with exactly the same thermal noise, i.e. using the same random number sequence for the three replicas. The spins in the replicas that begin with an ordered configuration will only flip due to thermal noise. As a consequence, a simultaneous flip in all three replicas is considered as due to thermal noise. On the other hand, if a spin does not flip simultaneously in all three replicas it is due to a domain wall.

\begin{figure}[!tbp]
\includegraphics[width=\columnwidth,clip=true]{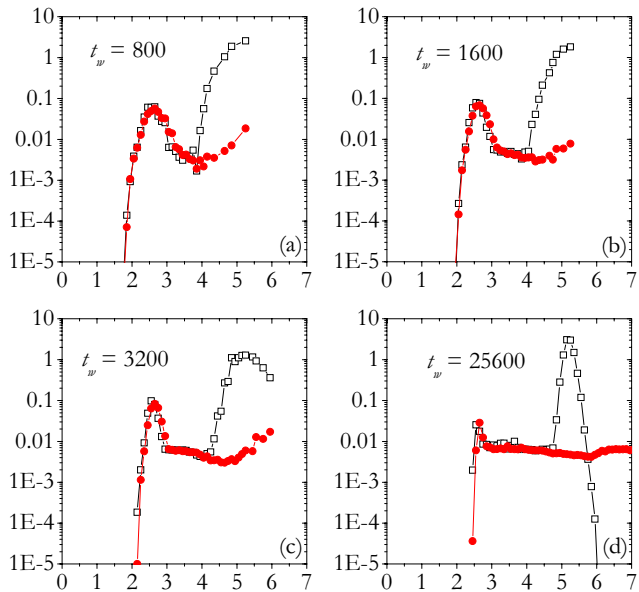}
\caption{\label{figure6} (Color online) Mean flipping time distribution $P$ vs. $\ln \tau_{\text{F}}$ (open square) and mean domain-wall-time distribution $Q$ vs. $\ln \tau_{\text{W}}$ (solid circle), of the 3D Ising model at $T=2.0$ for different values of $t_{\text{w}}$ with $\Delta t = t_{\text{w}}$. The average were carried out over $M=5\times10^4$ for $L=10$.}
\end{figure}

We can thus determine $N_{\text{W}}$, the number of time steps that a spin spends in a domain wall within the time window $\Delta t$. Thus we can compute the distribution function of the mean time $\tau_{\text{W}} = \Delta t / N_{\text{W}}$ that a spin spends in a domain wall, i.e. the mean domain-wall-time distribution $Q(\ln \tau_{\text{W}})$. Note that this is also an out-of-equilibrium quantity depending on $t$ and $t_{\text{w}}$. Figure~\ref{figure6} shows the evolution of $Q(\ln \tau_{\text{W}})$ at different times, and also a comparison with $P(\ln \tau_{\text{F}})$. As can be observed, $Q$ is also a highly heterogeneous function. For short time scales the left peak of both distributions coincide. For increasing times the valley in $Q$ becomes broader, and eventually the distribution becomes almost homogeneous beyond the first peak, which is always present. This behavior is somehow mirrored by $P$, that, as already described above, presents a decay in the fast peak.

The physical interpretation of the behaviors of $P$ and $Q$ is intimately related to the domain growth process in the Ising model. For low temperatures most flips will be due to domain walls, while flips inside a domain will take place with a very low probability. For short times the domains are very small, and the dynamics is fast in the sense that there are many flips due to domain walls dynamics, and both distributions coincide. For increasing time, domains grow and the fraction of spins in domain walls decay. This is reflected in the behavior of both $P$ and $Q$ for small values of, respectively, $\tau_{\text{F}}$ and $\tau_{\text{W}}$, and establishes a relation between dynamical heterogeneities and domain growth in the Ising model. Therefore, the strong correlation between $P$ and $Q$ observed in Fig.~\ref{figure6} indicates that the characteristic decay with time of the first peak is a hallmark of a domain growth process.

The first question that comes immediately to mind is if it is possible to relate some kind of phase ordering to the dynamical heterogeneities observed in the 3D EA model as we just did for the Ising model. In the following sections we will tackle this issue.

\section{Gauge transformation}
\label{sec:trans}

\begin{figure}[!tbp]
\includegraphics[width=4cm,clip=true]{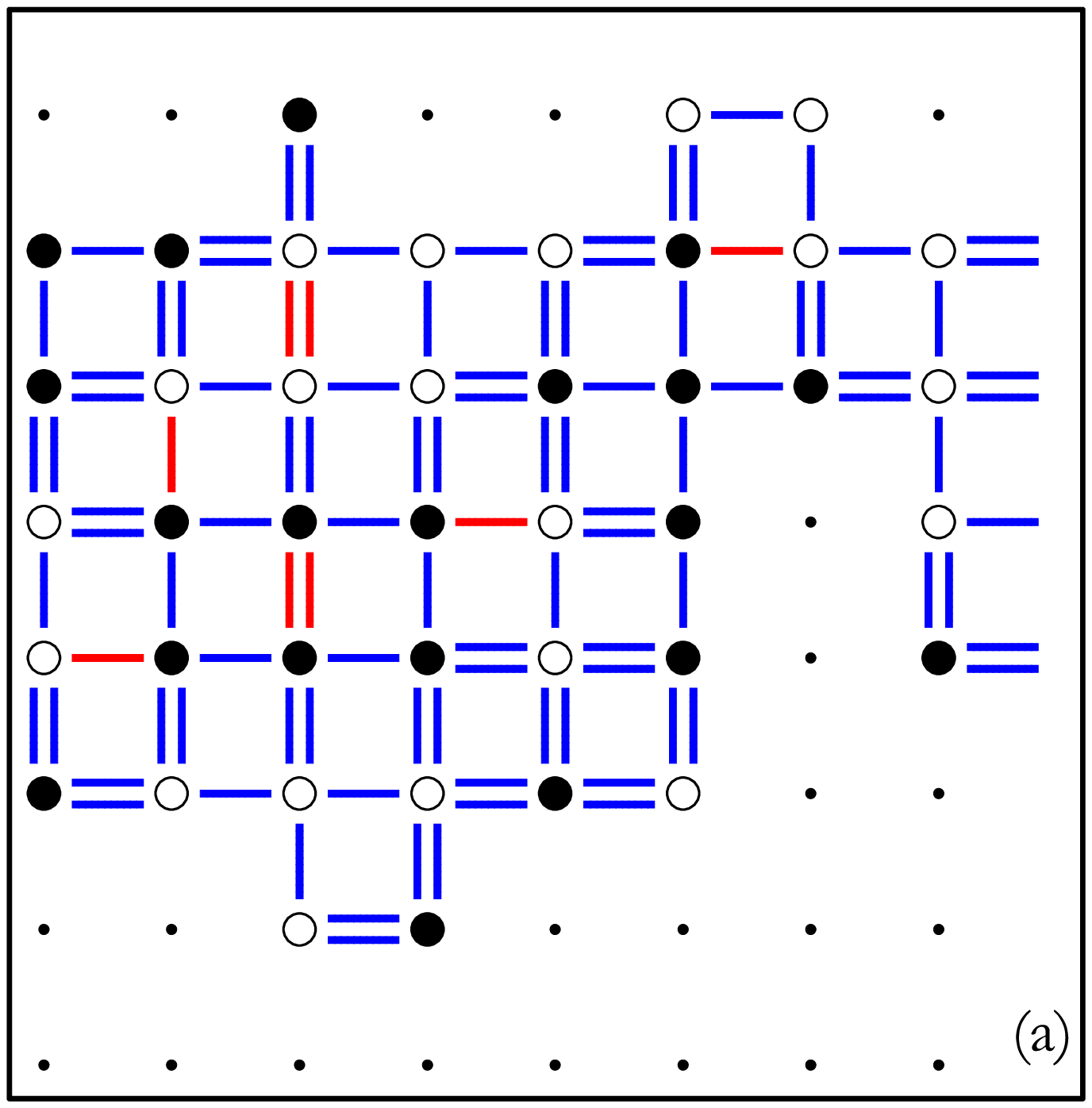}
\includegraphics[width=4cm,clip=true]{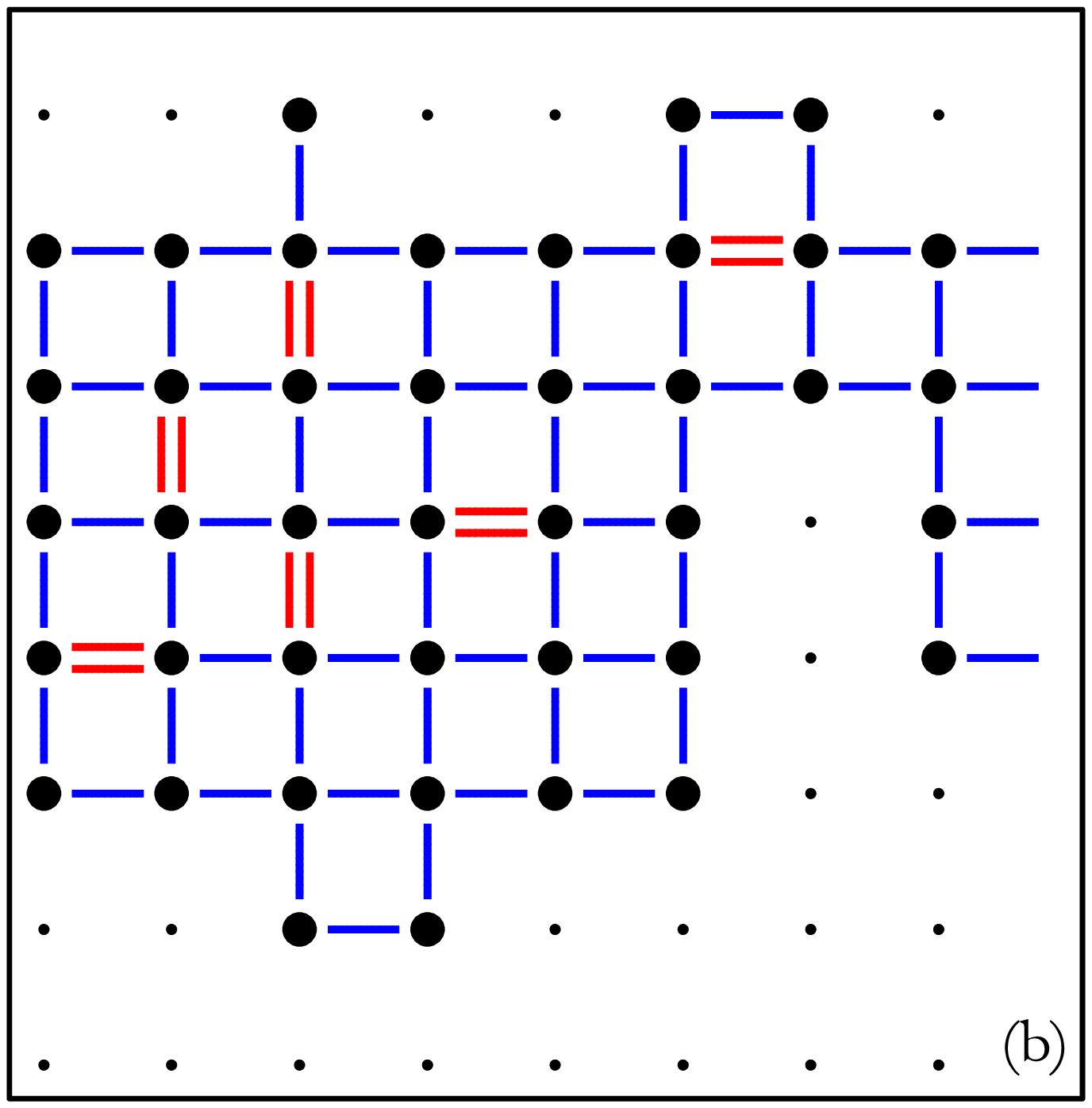}
\caption{\label{figure7} (Color online) Two-dimensional example of the gauge transformation. (a) The GS configuration of solidary spins: black (white) dots represents up (down) spins, single (double) lines correspond to ferromagnetic (antiferromagnetic) bonds, and red (blue) color corresponds to frustrated (satisfied) bonds. (b) After the gauge transformation all spins are up, all ferromagnetic bonds are not frustrated and all antiferromagnetic bonds.}
\end{figure}

In this section we shall present evidence that supports the idea that the backbone can sustain a ferromagnetic-like ordered structure.~\cite{Roma2009a} Recall that in Sec.~\ref{sec:model} we pointed out that in three dimensions the backbone percolates. Then, here we will focus our interest by restricting the following analysis to the largest cluster of solidary spins. Thus, we consider the Hamiltonian
\begin{equation}
 H_{\text{s}} = \sum_{( i,j ) \in \{s\}} J_{ij} \sigma_{i} \sigma_{j},
\label{hamiltonian-sol}
\end{equation}
where the sum has been restricted to this percolating set of solidary spins. Then, we consider the new spin variables defined through $s_i = \sigma_i^0 \sigma_i$, where $\{\sigma_i^0\}$ represent one of the two possible values of the spins variables in all the GS configurations. Now, by replacing the bonds by $J'_{ij} = J_{ij} \sigma_i^0 \sigma_j^0$, and since $(\sigma_i^0)^2=1$ one arrives at the Hamiltonian
\begin{equation}
 H'_{\mathrm{s}} = \sum_{( i,j ) \in \{\mathrm{s}\}} J'_{ij} s_{i} s_{j},
\label{hamiltonian-sol-gauge}
\end{equation}
which is indeed equivalent to $H_{\mathrm{s}}$. Therefore this gauge transformation leaves the Hamiltonian invariant. It is clear that in the GS the new spin variables are all $s_i^0= +1$, or equivalently the configuration with the opposite sign.

As an illustrative example, consider the set of solidary spins in Fig.~\ref{figure7}(a), which corresponds to one of the two possible GS configurations of the largest cluster of the backbone in a particular realization of a two-dimensional system. In the figure, single (double) lines correspond to ferromagnetic (antiferromagnetic) bonds from the bimodal distribution; red or blue color indicates if the interaction is frustrated or not, respectively. Figure~\ref{figure7}(b) represents the new set of spins $\{s_i^0\}$ with the corresponding bonds $J'_{ij}$. Notice that after the transformation all $s_i^0$ spins have the same sign. Another property of the gauge transformation is that it preserves the frustration in the system. Note that all the frustrated (red) bonds in Fig.~\ref{figure7} are always frustrated. Indeed, in three dimensions approximately only a $10\%$ fraction of the bonds linking spins within the backbone are frustrated. In this way we have obtained, after the gauge transformation, a ferromagnetic-like system with a $10\%$ fraction of frustrated antiferromagnetic bonds. Since $0.1 < x_c=0.222$, the limit of concentration of antiferromagnetic bonds in the ferromagnetic phase,~\cite{hartmann1999} \textit{and} since the backbone structure percolates in three dimensions,~\cite{Roma2009a} we expect that the set of solidary spins is a good candidate for ferromagnetic-like order. In particular, we expect that the qualitative behavior of the evolution of $P$ and $Q$ for solidary spins will somehow resemble the ones observed in the Ising model.

\section{Domain growth within the backbone}
\label{sec:growth}

\begin{figure}[!tbp]
\includegraphics[width=\columnwidth,clip=true]{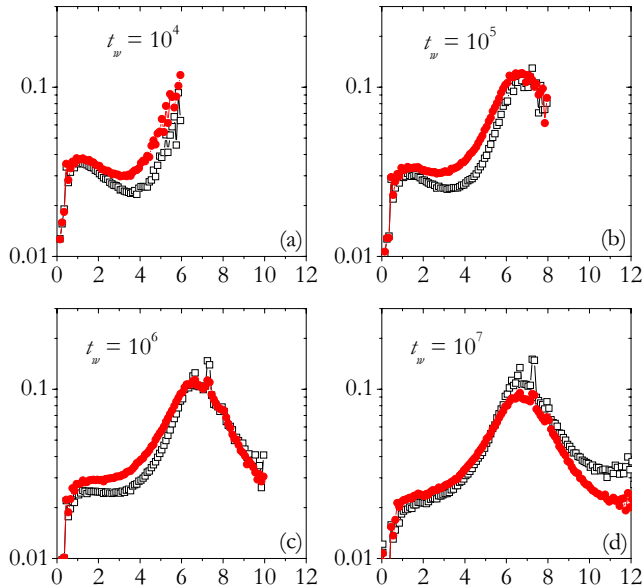}
\caption{\label{figure8} (Color online) The distributions $P_\text{s}$ vs. $\ln \tau_{\text{F}}$ (open square) and $Q_\text{s}$ vs. $\ln \tau_{\text{W}}$ (solid circle) corresponding only to the subset of solidary spins of the EA model. Data correspond to $T=0.5$ for different values of $t_{\text{w}}$ with $\Delta t = t_{\text{w}}$. The average were carried out over $M=10^3$ and $m=10$
for $L=8$.}
\end{figure}

In this section we shall present further numerical evidence for a growing ferromagnetic-like process in the EA model. In Sec.~\ref{sec:ising} we analyzed the mean flipping time distribution for the Ising model. We observed that this distribution presents a bimodal shape, and we were able to establish a direct relation between the time that spins spend in a domain wall and the changes in shape that takes place in the mean flipping time distribution as a domain growth process takes place.

In order to establish a direct comparison with the results of the previous sections we will present a generalization of the algorithm for domain wall detection of Hinrichsen and Antoni,~\cite{Hinrichsen1998} originally devised for the Ising model. Since the $\pm J$ EA model has a highly degenerate GS, the perspective of simulating the system and replicas in all the GS configurations seems a daunting task. However, as has been stressed through the article, the ground state of the 3D EA model is very peculiar, and dictates how the systems can be decomposed in solidary and non-solidary spins. Since, by definition, solidary spins maintain their relative orientation in all the ground state configurations we can choose just one GS configuration --and the one with all the spins reversed-- to compare with the replica that has a random initial condition. In this way the algorithm will permit us to identify domain walls {\em within} the backbone, which is the structure with a simple symmetry break. On the other hand, we will not be able to obtain any relevant information outside the backbone. Once the domain walls have been identified one can compute the mean domain-wall-time distribution, $Q$.~\footnote{We have compared the results obtained using one GS configuration with the one obtained by averaging over many GS configurations and we checked that there are not changes.}

In Fig.~\ref{figure8} we present the mean flipping time distribution for solidary spins $P_{\text{s}}$ vs. $\ln \tau_{\text{F}}$ (open square), together with the mean domain-wall-time distribution for solidary spins $Q_{\text{s}}$ vs. $\ln \tau_{\text{W}}$ (solid circle) for $T=0.5$. As can be observed, both distributions are highly correlated, presenting  similar qualitative behaviors. The peak on the right does not seem to depend on the waiting time. On the other hand, for short waiting times the distributions present a peak on the left, which decreases with increasing waiting time, and eventually disappears. Therefore, the mean flipping time and the mean domain-wall-time distributions for solidary spins are highly correlated and presents a first peak which displays a fast decay with increasing waiting time.  As was observed for the Ising model in Sec.~\ref{sec:ising}, this behavior is intimately related to a domain growth proccess. Thus, the comparison with the Ising model and also the small bond frustration present on the backbone suggest a plausible scenario for ferromagnetic-like growing order in the 3D $\pm J$ EA model.

\section{Discussions and conclusion}
\label{sec:conc}

Throughout the present work, based on the separation in solidary and non-solidary spins given by the GS topology, we have presented evidence supporting a picture of growing ferromagnetic-like order within the backbone of the 3D $\pm J$ EA model.

By studying the two-times correlation function we showed that the solidary/non-solidary spin separation is not trivial in the sense that it is able to give meaningful dynamical information. While solidary spins maintain their correlation in time, non-solidary spins present a fast initial decay of the correlation function.

On the other hand, the mean flipping time distribution shows that solidary (non-solidary) spins mainly contribute to the slow (fast) peak. Besides, although it can be shown that a given spin contributes to both peaks, the time evolution of the mean flipping time distribution for solidary and non-solidary spins behaves qualitatively different. As a main characteristic, the left peak of the solidary spins $P_{\text{s}}$ rapidly decreases with increasing waiting time. In order to correlate this information with ferromagnetic order we have also analyzed the pure ferromagnetic Ising model using the same protocol. Thus, we were able to associate the heterogeneous behavior of $P(\ln \tau_{\text{F}})$ with the mean domain-wall-time distribution $Q(\ln \tau_{\text{W}})$. The results obtained are the expected for a coarsening system like the Ising model and give us a quantitative tool to analyze the mean flipping time distribution in the EA model.

Seeking for ferromagnetic order in the 3D $\pm J$ EA spin glass model, we showed that using a gauge transformation the solidary spins can be transformed into a ferromagnetic system with a small fraction of frustrated antiferromagnetic bonds. With this in mind we presented a generalization of the algorithm for domain wall detection of Hinrichsen and Antoni. This allowed us to present a quantitative characterization of $P_{\text{s}}(\tau_{\text{F}})$ and $Q_{\text{s}}(\tau_{\text{W}})$ and, based in the comparison with the Ising model, we conclude that ferromagnetic-like order is growing within the backbone of the model.

All these results, together with other recent works that show the relevance of the heterogeneous character of the GS topology,~\cite{Roma2,Roma3,rubio2009,Roma2007} point to an alternative picture for the nature of the spin glass phase. Within this picture, ferromagnetic-like order grows within a constrained structure of the system. It was precisely with this idea in mind that we were able to present a generalization of the method proposed by Hinrichsen and Antoni, and thus characterize a ferromagnetic-like domain growth process.

Finally, we stress the non-trivial character of the results by comparing with the Gaussian spin glass model that has only one simple degenerated GS. Note that even when a gauge transformation can lead to a disorder ferromagnetic system, the generalization of the domain wall detection method is not direct due to the high frustration of the whole system. In order to actually obtain relevant physical information it would be thus necessary to determine some kind of low frustrated constrained structure. Concluding, the results presented in this work highlight the importance of the determination of the backbone structure in spin glass models.

\begin{acknowledgments}
The authors thank fruitful discussions with J. Arenzon, S. Cannas, M. L. Rubio Puzzo, D. Stariolo, and F. Tamarit. FR acknowledges financial support from  projects PICT05-33328 and PICT07-2185, and U.N. de San Luis under project 322000.
\end{acknowledgments}

\appendix*
\section{Finite size effects}
\label{sec:app}

\begin{figure}[!tbp]
\includegraphics[width=7cm,clip=true]{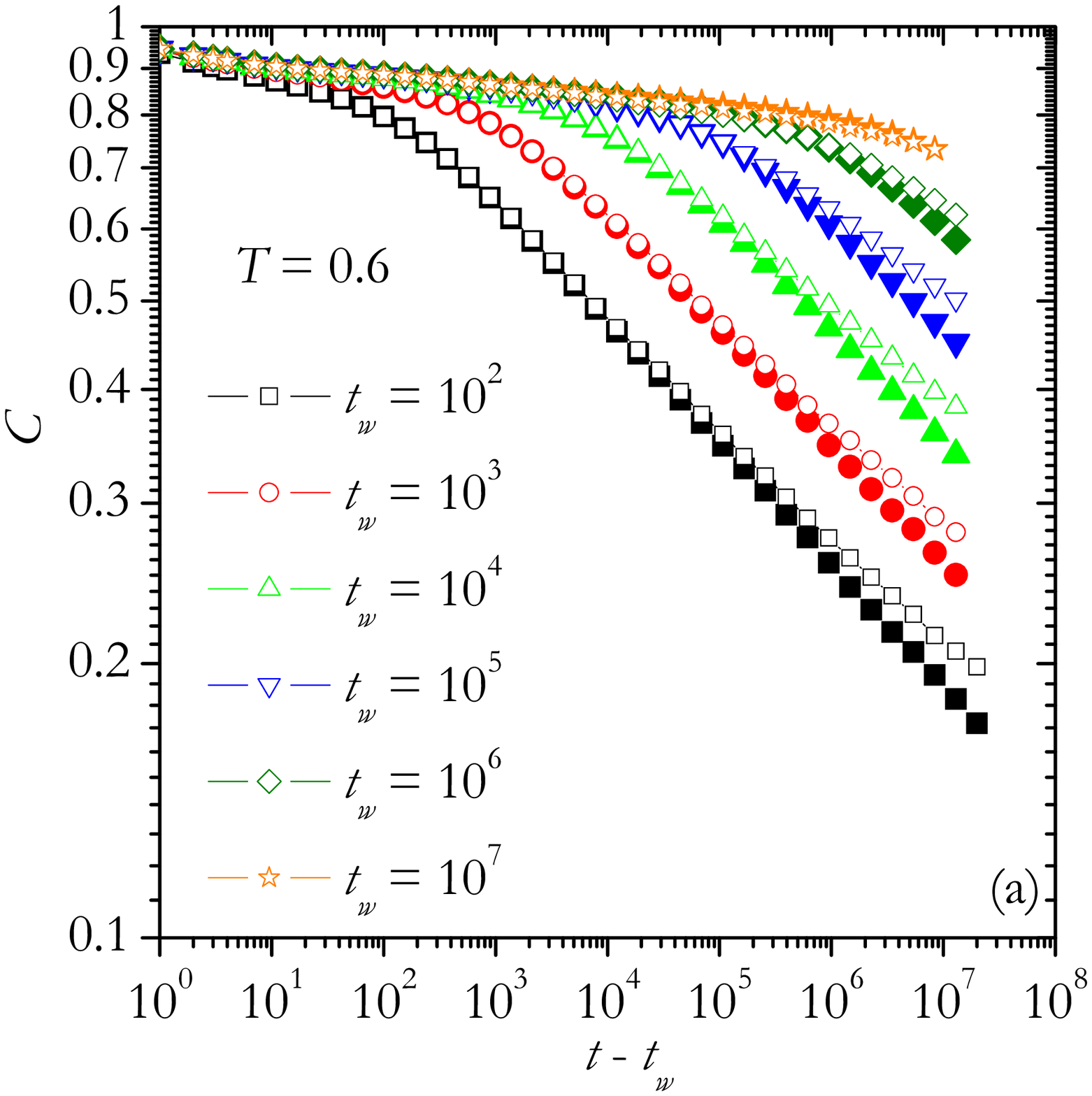}
\includegraphics[width=7cm,clip=true]{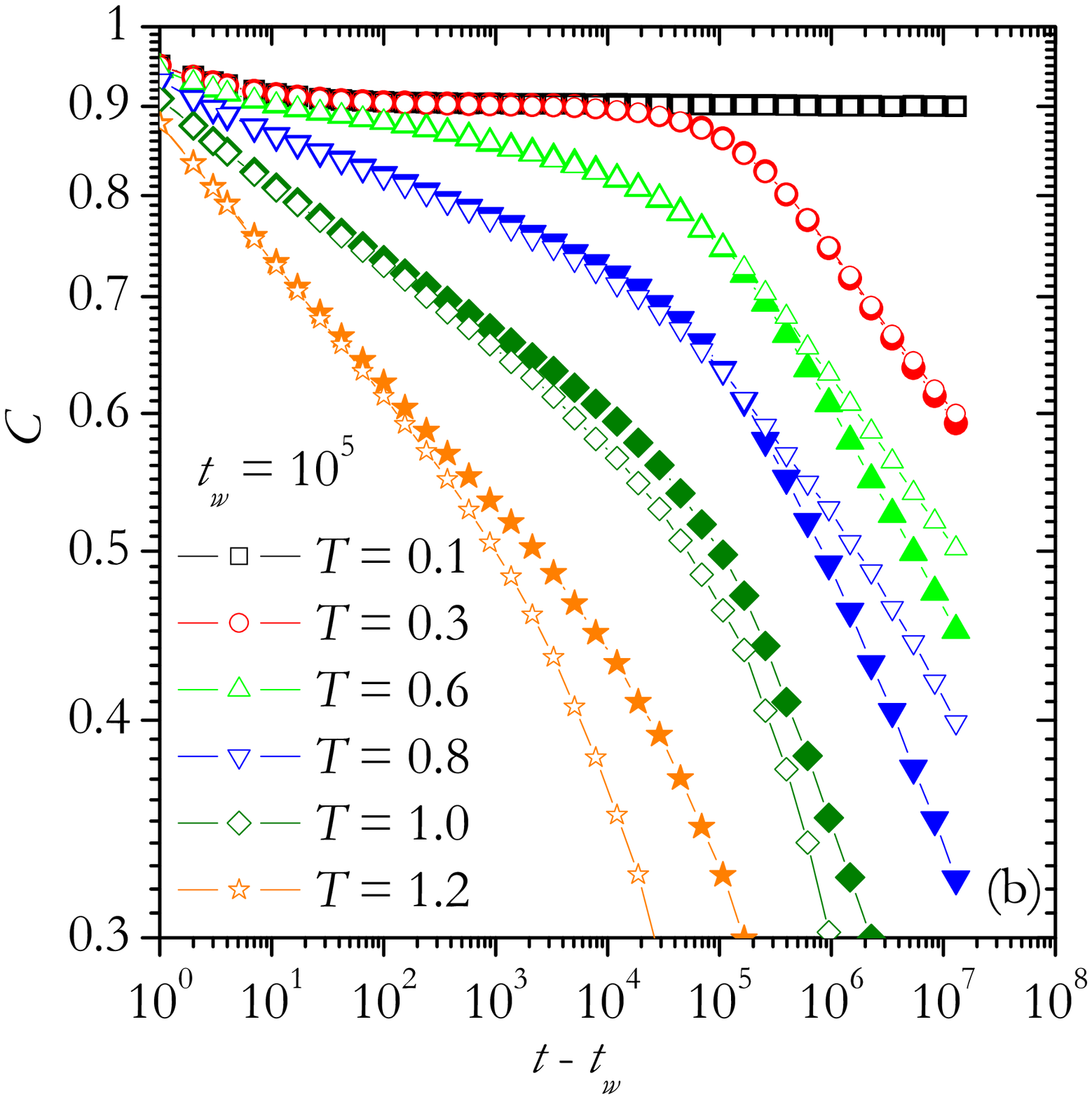}
\caption{\label{figure9} (Color online) Two-times correlation function for $L=8$ (open symbols) and $L=20$ (solid symbols) systems at (a) $T=0.6$ for different values of $t_{\text{w}}$, and at (b) $t_{\text{w}}=10^5$ for different temperatures as indicated. The averages were carried out over $M=10^3$ ($L=8$) and $M=10^2$ ($L=20$) samples, with $m=10$ thermal histories for each sample.}
\end{figure}

\begin{figure}[!tbp]
\includegraphics[width=7cm,clip=true]{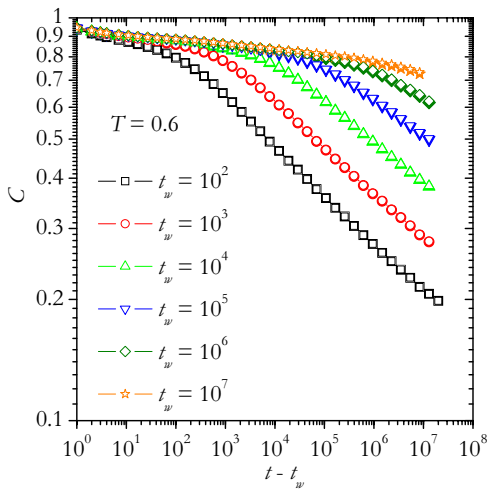}
\caption{\label{figure10} (Color online) Correlation function for
$L=8$ systems at $T=0.6$, calculated with Eq.~\eqref{corr} (open
symbols) and Eq.~\eqref{suma_corr} (solid symbols). The average
were carried out over $M=10^3$.}
\end{figure}

We shall show within this appendix that finite size effects are not strong in the time window we are interested in through the paper. For this purpose, although one can find the temperature and time dependence of the two-times correlation function in previous works,~\cite{rieger1993,jimenez2003} in the following we use the out-of-equilibrium correlation function to carefully compare different system sizes. Figure~\ref{figure9}(a) shows $C(t,t_{\text{w}})$ for a fixed temperature $T=0.6 < T_{\mathrm{g}}$ and different waiting times $t_{\text{w}}$. Data are shown for two system sizes, $L=8$ and $L=20$. Up to delay-times of the order of $t-t_{\text{w}} \sim 10^5$ the curves overlap. For larger delay-times small deviations are observed due to finite size effects. However, we stress that the same qualitative behavior is observed for both system sizes. The correlation function for the same system sizes is also considered in Fig.~\ref{figure9}(b), but as a function of temperature for fixed $t_{\text{w}}=10^5$. It is observed that finite size effects are stronger at higher temperatures. However, these are not significant for temperatures below $T=0.6$ and times such that $\Delta t=t_{\text{w}}=10^5$. For small temperatures and larger times there are small deviations between the $L=20$ and $L=8$ curves, which nevertheless show a good qualitative agreement.

A different approach we use to quantify finite size effect comes from the information obtained with the separation of the correlation function in its solidary and non-solidary contributions, as explained in Sec.~\ref{sec:corr}. Defining the mean fractions of solidary and non-solidary spins as, respectively, $f_\text{s}=\left[ N_\text{s}^{\alpha} / N \right ]_{\text{av}}$ and $f_\text{n}=\left [ N_\text{n}^{\alpha} / N \right]_{\text{av}}$ with $f_\text{s} + f_\text{n} = 1$, then in the thermodynamic limit it should hold that
\begin{equation}
C(t,t_{\text{w}})=f_\text{s} C_\text{s}(t,t_{\text{w}}) + f_\text{n}
C_\text{n}(t,t_{\text{w}}).  \label{suma_corr}
\end{equation}
This would be strictly valid in the thermodynamic limit. If finite size effects are important, the mean fractions would present important fluctuations and Eq.~\eqref{corr} could not be split in the solidary and non-solidary contributions. We show in Fig.~\ref{figure10} that, although the mean fractions do fluctuate,~\cite{Roma2009a} the total correlation function is very well approximated by Eq.~\eqref{suma_corr}, even for $L=8$. Data correspond to the full correlation function computed using Eq.~\eqref{corr} (open symbols) and the one computed from Eq.~\eqref{suma_corr} (solid symbols). Notice that the different data sets are almost indistinguishable, even for the small system size we are using. In this case, averages were taken over $M=10^3$ independent samples which yields $f_\text{s}=0.75883$ and $f_\text{n}=0.24117$.

Summarizing, in this appendix we have performed a finite size study by comparing two system sizes, $L=8$ and $L=20$. We have shown that for the low temperatures and for the time scales we are interested in, physically relevant information can be extracted from a system of linear size $L=8$. Thus, to analyze two-times dependent observables of the 3D $\pm J$ EA model we used this system size, where we have determined the backbone structure for an extensive number ($M=1000$) of disorder realizations.

\bibliography{spinglass,structural-glass}
\bibliographystyle{apsrev}

\end{document}